\documentstyle[twoside,fleqn,espcrc2,amsfonts]{article}

\newcommand{\AmS}{{\protect\the\textfont2
  A\kern-.1667em\lower.5ex\hbox{M}\kern-.125emS}}
\newcommand{\psibar}{\overline{\psi}}
\newcommand{\kH}{{\cal{H}}}
\def\identity{{\Bbb I}}                      
\title{Ginsparg--Wilson Relation and Spin Chains}

\author{Ivan Horv\'ath\address{Department of Physics, University of Virginia,
        Charlottesville, VA 22903, USA} 
        \thanks{The speaker. Work supported in part by the DOE under
        grant No. DE-FG02-97ER41027.}
        and 
        Harry~B. Thacker${}^{\rm a}$}
       
\begin{document}

\begin{abstract}
The Baxter 8-vertex model is equivalent to a particular lattice 
formulation of a self-interacting, massive Dirac fermion theory.
In the time-continuum limit, the lattice Hamiltonian (XYZ spin
chain) can be explicitly transformed to a lattice Dirac Hamiltonian.
We show that the kernel describing the quadratic part of this
Hamiltonian satisfies a one-dimensional version of the Ginsparg-Wilson
relation. The corresponding conserved charge is derived and compared with
the conserved arrow number of the 8-vertex model.

\end{abstract}

\maketitle

The continuum field theory obtained at the second-order phase transition
of the 8-vertex
model is equivalent to the massive Thirring model, a theory 
of a self-interacting Dirac fermion in two space-time 
dimensions~\cite{Lut75A}. 
This was demonstrated by examining the continuum
limit of the $XYZ$
nearest-neighbor Heisenberg spin-chain Hamiltonian, which is
obtained from the 8-vertex model transfer matrix 
in the infinite anisotropy (i.e. time-continuum) limit.
At the point where the mass is zero, the spin model reduces to a 6-vertex
model (or the $XXZ$ spin chain in the Hamiltonian limit)
which exhibits a conservation of arrows at each vertex.
In the spin chain Hamiltonian, this arrow conservation
reduces to the conservation of the $z$-component of the 
total spin, and   
is a consequence of the symmetry under a global rotation
of spins in the $\sigma^x-\sigma^y$ plane. Some time ago,
it was argued that the conserved arrow number of the spin
chain Hamiltonian corresponded to an exact lattice chiral symmetry
of the equivalent fermion theory~\cite{Tha95A}. This argument
relied on the transformation properties of the lattice
fermion under an exact lattice Lorentz invariance~\cite{Tha86A}.

Recent developments on the problem of chiral lattice
fermions~\cite{Has98B,Neu98A,Lus98A} have focused on the 
Ginsparg-Wilson (GW) relation\cite{Gin82A}.
If a lattice theory is defined by a kernel that satisfies this relation,
then it has an analog of exact chiral symmetry~\cite{Lus98A}. 
The Nielsen-Ninomiya theorem is avoided by the fact that the chiral 
transformation is not a simple on-site $\gamma^5$ rotation, but involves
hopping terms present in the corresponding Dirac kernel. Here we will
investigate the chiral symmetry of the 8-vertex model in the light of
these new developments.

In the massless vertex model the conserved arrow charge is
associated with the symmetry of the vertex Boltzmann weight
under a local phase rotation of the four arrows involved in a
single vertex. However, as we show here, the lattice Dirac spinor 
of the equivalent fermion Hamiltonian is constructed from
combinations of fermionized spins of the $XYZ$ chain, residing
on different sites. Thus, the on-site phase rotation of the arrows
becomes a transformation which mixes Dirac components on neighboring sites.
This raises the possibility that the chiral symmetry of the
vertex model is realized in a Ginsparg-Wilson form and that 
the conserved arrow charge is related to the charge constructed
by L\"uscher for theories satisfying the GW relation. Here
we present some evidence to support this proposition. Since
a direct transformation from the 2-dimensional vertex model
to the 2-dimensional Dirac action has not been constructed yet,
our analysis focuses on the symmetry of the massless spin chain. 

In the present discussion, we will consider only the free fermion
part of the Hamiltonian (the $XY$ chain). After performing a Jordan-Wigner
transformation which transforms the spins into canonical fermion operators, 
the Hamiltonian may be written as
\begin{equation}
\label{eq:XY}
   H = i\sum_j c^x_{j+1} c^y_j + k c^x_j c^y_{j+1}\;,
   \label{eq:5}
\end{equation}
where the operators $c^x_j,c^y_j$ satisfy
\begin{equation}
   (c^a_j)^\dagger = c^a_j \quad\; 
   \{c^a_j,c^b_k\} = \delta_{ab}\delta_{jk}\quad\;
   a,b=x,y\;.
\end{equation}	
The massless limit of (\ref{eq:XY}) is $k\rightarrow 1$.
Our aim is to put the above quadratic Hamiltonian in the standard form,
involving a complex Dirac spinor. Since there are two real degrees
of freedom per site, we can combine them to define a single complex
canonical variable per site, namely
\begin{eqnarray}
   c_j &=& {1\over \sqrt{2}}(c^x_j + i c^x_{j+1})\qquad j\; \mbox{even}  \nonumber \\
   c_j &=& {1\over \sqrt{2}}(c^y_j + i c^y_{j+1})\qquad j\; \mbox{odd}\,.  
   \label{eq:10}
\end{eqnarray}
By construction, these variables are canonical and satisfy
\begin{equation}
   \{c_j^\dagger,c_k\} = \delta_{jk}\;,
\end{equation}
with all other anticommutators vanishing. Defining relations (\ref{eq:10}) can be easily
inverted and the Hamiltonian takes the form
\begin{eqnarray}
   H &=& i\sum_n c_{2n} c_{2n-1}^\dagger + c_{2n}^\dagger c_{2n-1} \nonumber \\
     &+& k\,(\,c_{2n} c_{2n+1}^\dagger + c_{2n}^\dagger c_{2n+1})\,.
   \label{eq:15}
\end{eqnarray}

	Next, we define a Dirac spinor $\psi_n$ living on the sublattice of the
original lattice through
\begin{equation}
   \psi_n^1 = (-1)^n c_{2n-1}\qquad \psi_n^2 = (-1)^n c_{2n}\,,
   \label{eq:17}
\end{equation}
where the factors $(-1)^n$ were introduced for later convenience.
The resulting Hamiltonian on the decimated lattice then reads
\begin{eqnarray}
   H &=& i\sum_n\, (\psi_n^2)^\dagger \psi_n^1 -
                 (\psi_n^1)^\dagger \psi_n^2       \nonumber \\
     &-&   k\,\Bigl(\,(\psi_n^2)^\dagger \psi_{n+1}^1 -
                 (\psi_{n+1}^1)^\dagger \psi_n^2\,\Bigr) \;.
   \label{eq:20}
\end{eqnarray}
Note that we could have switched the ``mass'' and ``hopping'' terms of the above 
Hamiltonian by shifting the indices in defining relations (\ref{eq:17}) by one. 
  
	Finally, we introduce the Dirac structure by defining
\begin{equation}
\gamma_0=\pmatrix{0&-i\cr
                  i&0\cr} \qquad
\gamma_1=\pmatrix{-1&0\cr
                  0&1\cr}\,.
\end{equation}
This brings the Hamiltonian to the form 
\begin{eqnarray}
   H = \sum_n \psibar_n\psi_n
    &+& {k\over 2}\sum_n \psibar_n (\gamma_1-1)\psi_{n+1} \nonumber \\
    &-& {k\over 2}\sum_n \psibar_n (\gamma_1+1)\psi_{n-1}\,,
   \label{eq:25}
\end{eqnarray}
where $\psibar=\psi^\dagger\gamma_0$. Comparing this to the standard form of the
Wilson Hamiltonian $H_W(K,M,r)$
\begin{eqnarray}
  H_W &=& M\sum_n \psibar_n\psi_n +
             K\sum_n \psibar_n (\gamma_1-r)\psi_{n+1} \nonumber \\
             &+& K\sum_n \psibar_n (\gamma_1+r)\psi_{n-1}\,,
\end{eqnarray}
we conclude that the Hamiltonian (\ref{eq:5}) is equivalent to $H_W(k/2,1,1)$,
and describes a free Wilson fermion with a specific choice of Wilson parameters.

      To see the connection to the GW condition and corresponding symmetry, it
is convenient to work in momentum space which we define by
\begin{equation}
    \psi_n = {1\over {2\pi}}\int_{-\pi}^{\pi}dp\,\psi(p)e^{ipn}\,.
\end{equation}
If we represent the Hamiltonian (\ref{eq:25}) by kernel $\kH$ through
$H=\psibar_m\kH_{mn}\psi_n$, then in Fourier space we have
$H = 1/2\pi\int_{-\pi}^{\pi}dp\,\psibar(p)\kH(p)\psi(p)$, where
\begin{equation}
   \kH(p) = 1 - k\cos(p) + ik\sin(p)\,\gamma_1\,.
\end{equation}
The one--particle spectrum of this theory can be found by diagonalizing
$\gamma_0\kH(p)$ and is given by
\begin{equation}
  \epsilon(p)^2 = (1-k)^2 + 2k\bigl(1-\cos(p)\bigr)\,.
\end{equation}
Consequently, the spin chain Hamiltonian (\ref{eq:5}) describes a massless
relativistic fermion if $k=1$. Restricting ourselves to that value and 
denoting the corresponding kernel by $\kH^c$, it is easy to check that
\begin{equation}
   \kH^c\gamma_5 + \gamma_5\kH^c = \kH^c\gamma_5\kH^c\,.
\end{equation}   
In other words, $\kH^c$ satisfies the Ginsparg-Wilson relation.

While the above observation is quite amusing, it is not at all obvious that the
Ginsparg--Wilson condition for the Hamiltonian kernel actually has any interesting
symmetry consequences for the theory as it does in the case of the Euclidean 
formulation~\cite{Lus98A}. To investigate this, consider some generic quadratic
Hamiltonian $\psibar\kH\psi$, on the odd--dimensional spatial lattice, such that
the kernel $\kH$ satisfies GW relation. Let us consider the quantity
\begin{equation}
\label{eq:Q}
   Q = \psibar\,\gamma_0\gamma_5 (\,1-{1\over 2} \kH\,)\,\psi\,,
\end{equation}
which is constructed in analogy to the one involved in the chiral transformation
considered by L\"uscher~\cite{Lus98A}. In the continuum limit it reduces to the 
standard axial charge. If we require $Q$ to be conserved, i.e.
\begin{equation}
  [\,\gamma_0\kH,\,\gamma_5(1-{1\over 2} \kH)\,]=0 \,, 
\end{equation}
then, in addition to GW condition, we must also impose
\begin{equation}
  [\,\kH, \gamma_0\gamma_5\,] = 0 \,.
  \label{eq:gw-extra}         
\end{equation}
In one dimension, this additional condition reduces to $[\,\kH,\gamma_1\,]=0$, which is 
fulfilled for our $\kH^c$, and $Q$ is thus indeed conserved in this case.

It is interesting to note that using the methods of Ref.~\cite{Hor98A} it can be
shown that $\kH^c$ is the only acceptable ultralocal solution of the GW relation
in one spatial dimension, satisfying condition (\ref{eq:gw-extra}). Indeed,
taking (\ref{eq:gw-extra}) into account, the most general kernel $\kH$ can be
written in form
\begin{equation}
   \kH(p) = \bigl(1-A(p)\bigr)\identity + iB(p)\gamma_1\,.
\end{equation}  
The GW relation then translates into the algebraic condition $A^2+B^2=1$. 
Using a basic Lemma proved in Ref.~\cite{Hor98A}, it then follows that a unique 
ultralocal solution giving massless relativistic spectrum is 
$A(p)=\cos(p),\, B(p)=\sin(p)$. This corresponds to $\kH^c(p)$. No ultralocal 
solutions in higher dimensions respecting hypercubic symmetry exist.

To see the relation of $Q$ to the conserved arrow charge of the vertex model,
it is useful to write it in terms of the spin-chain fermion operators, namely
\begin{equation}
Q = {i\over 2} \sum_j c^x_jc^y_j + 
    c^y_{2j+2}c^x_{2j} + c^y_{2j-1}c^x_{2j+1}.
\end{equation} 
The ``on-site'' part is (up to proportionality constant) the arrow charge $Q_A$,
and we denote the ``next-nearest neighbour'' part as $Q_2$, i.e. $Q=Q_A+Q_2$. 
$Q_2$ is one of the higher conservation laws that exist in this model
as a manifestation of complete integrability. The relation between $Q_A$ and $Q$ 
is somewhat reminiscent of the relation between the Hamiltonian and the
$\log$ of the transfer matrix. The latter reduces to the (nearest-neighbor) Hamiltonian
in the time continuum limit, but for finite lattice spacing in the time direction, the
$\log$ of the transfer matrix includes higher conserved operators with higher hopping
terms. It would be useful to explore these connections in the context of the full
2-dimensional vertex model, rather than being restricted to the Hamiltonian limit.
An explicit construction of the 2-dimensional lattice Dirac operator ${\cal D}$ for
this model would be of great interest. Heuristic arguments suggest that
${\cal D} $ will not be ultralocal, in accordance with the no-go theorem of 
Ref.~\cite{Hor98A}, and so the possibility that ${\cal D}$ satisfies the 2-dimensional
Ginsparg-Wilson relation is not ruled out.

Let us finally note that kernel like $\kH^c$ was actually considered in the context of
perfect fermionic actions~\cite{Bie94A}. Now, that the relation of perfect actions 
to Ginsparg-Wilson approach is clear, this is actually not surprising. We thank 
S.~Chandrashekharan, P.~Hasenfratz and W.~Bietenholz for pointing that out to us.


\begin{thebibliography}{9}
\bibitem{Lut75A} A.~Luther, Phys.~Rev. {\bf B14} (1975) 2153.
\bibitem{Tha95A} H.B.~Thacker, Nucl.~Phys. {\bf B} (Proc. Suppl.) {\bf 42} (1995) 642.
\bibitem{Tha86A} H.B.~Thacker, Physica~{\bf 18D} (1986) 348.
\bibitem{Has98B} P.~Hasenfratz, hep-lat/9802007.
\bibitem{Neu98A} H.~Neuberger, Phys.~Lett. {\bf B427} (1998) 353.
\bibitem{Lus98A} M.~L\"{u}scher, Phys.~Lett. {\bf B428} (1998) 342.
\bibitem{Gin82A} P.~Ginsparg and K.~Wilson, Phys.~Rev. {\bf D25} (1982) 2649.
\bibitem{Hor98A} I.~Horv\'ath, hep-lat/9808002.
\bibitem{Bie94A} W.~Bietenholz and U.-J.~Wiese, Nucl. Phys. {\bf B} (Proc. Suppl.) 34 (1994) 516.


\end{thebibliography}
\end{document}